          [2001/04/25 1.1 (PWD)]
\documentclass[a4paper,twocolumn]{esapub} 
\usepackage{natbib}

\title{The INTEGRAL view of the Soft Gamma-ray Repeater SGR 1806-20}
\author[1]{D. G\"{o}tz}
\author[1]{S. Mereghetti}
\author[2,3]{I.F. Mirabel}
\author[4]{K. Hurley}
\author[5]{S. Brandt}
\author[5]{N. Lund}
\author[6]{P. Ubertini}
\author[6]{M. Del Santo}
\author[6]{E. Costa}
\author[6]{M. Feroci}
\author[7,15]{P. Kretschmar}
\author[8]{A. Castro-Tirado}
\author[9]{A. Gimenez}
\author[10]{J.-L. Atteia}
\author[11]{M. Boer}
\author[12]{T. Cline}
\author[13,14]{F. Frontera}
\author[14]{G. Pizzichini}
\author[15]{A. von Kienlin}
\author[16]{E. G{\" o}{\u g}{\" u}{\c s}}
\author[17]{C. Kouveliotou} 
\author[17]{M. Finger}
\author[18]{C. Thompson}
\author[19]{H. Pedersen}
\author[20]{M. van der Klis}

\affil[1]{Istituto di Astrofisica Spaziale e Fisica Cosmica (IASF) - CNR, Milano, Italy}
\affil[2]{Service d'Astrophysique, CEA/Saclay, Gif-sur-Yvette, France}
\affil[3]{Instituto de Astronomia y Fisica del Espacio / CONICET, Buenos Aires, Argentina}
\affil[4]{UC Berkeley Space Sciences Laboratory, Berkeley, USA}
\affil[5]{Danish Space Research Institute (DSRI), Copenhagen, Danemark}
\affil[6]{Istituto di Astrofisica Spaziale e Fisica Cosmica (IASF) - CNR, Roma, Italy}
\affil[7]{INTEGRAL Science Data Centre (ISDC), Versoix, Switzerland}
\affil[8]{Instituto de Astrof\'{\i}sica de Andaluc\'{\i}a (IAA-CSIC), Granada, Spain}
\affil[9]{Centro de Astrobiolog\'{\i}a (CSIC-INTA), Madrid, Spain}
\affil[10]{Laboratoire d'Astrophysique, Observatoire Midi-Pyrene\'{e}s, Toulouse, France}
\affil[11]{Centre d'Etude Spatiale des Rayonnements, CNRS/UPS, Toulouse, France}
\affil[12]{NASA Goddard Space Flight Center, Greenbelt, USA}
\affil[13]{Universit\`{a} degli Studi di Ferrara, Ferrara, Italy}
\affil[14]{Istituto di Astrofisica Spaziale e Fisica Cosmica (IASF) - CNR, Bologna, Italy}
\affil[15]{Max-Planck-Institut f\"{u}r extraterrestrische Physik (MPE), Garching, Germany}
\affil[16]{Sabanci University, Orhanli-Tuzla, Istanbul, Turkey.}
\affil[17]{NASA/Marshall Space Flight Center, National Space Science and Technology Center, Huntsville, USA}
\affil[18]{Canadian Institute for Theoretical Astrophysics, University of Toronto, Toronto, Canada}
\affil[19]{Astronomical Observatory, University of Copenhagen, Copenhagen, Denmark}
\affil[20]{Astronomical Institute, University of Amsterdam, Amsterdam, the Netherlands}

\def\src  {SGR~1806--20~}
\usepackage{epsfig}

\begin{document}

\keywords{Gamma Rays : bursts; pulsars: general; stars: individual (SGR 1806-20)}
\maketitle

\begin{abstract}
We present the results obtained by INTEGRAL on the Soft-Gamma Ray Repeater
\src. In particular we report on the temporal and spectral properties of the bursts 
detected  during a moderately active period of the source in September and October 
2003 and on the search for quiescent emission.
\end{abstract}

\section{Introduction}
Soft Gamma-ray Repeaters (SGRs) are a class of peculiar 
high-energy sources discovered 
through their recurrent emission of soft $\gamma$-ray bursts.
These bursts have typical durations of $\sim$0.1 s 
and luminosities in the range 10$^{39}$-10$^{42}$ ergs s$^{-1}$
(see \citet{hurleyrew} for a review of this class of objects). 
Occasionally, SGRs also emit giant bursts that last up to a
few hundred seconds and exhibit remarkable pulsations that
reveal their spin periods (e.g. \citet{mazets}, \citet{hurley1999}).

The bursting activity and the persistent 
emission observed in the $\sim$0.5-10 keV energy
range are generally explained
in the framework of the ``Magnetar'' model (see e.g. \citet{dt92}, \citet{pac92},
\citet{td95}), as caused by a highly magnetized ($B\sim$10$^{15}$ G) slowly 
rotating ($P\sim$ 5-8 s) neutron star. In this model magnetic
dissipation causes the neutron star crust to fracture. These
fractures generate sudden shifts in the magnetospheric footpoints, which
trigger the generation of Alfv\'{e}n pulses, which in turn accelerate
electrons above the pair-production threshold, resulting quickly in
an optically thick pair-photon plasma. The cooling of this plasma
generates the typical short bursts of soft $\gamma$-ray radiation. The longer
bursts are powered by magnetic reconnection, and involve the entire neutron
star magnetosphere.

SGR 1806--20 is one of the most active Soft Gamma-ray Repeaters.
Here we report new observations of this source obtained with the 
INTEGRAL satellite in September and October 2003 during a period of moderate 
bursting activity \citep{mereb,gotza,hurley,merec,gotzb}.
These data have two advantages compared to previous observations
in the soft $\gamma$-ray energy range of bursts from this source.
First, they have been obtained with an imaging instrument, thus we can
exclude that the bursts originate from a different source in the field.
Second, they have a good sensitivity and time resolution which allows us 
to study the spectral evolution of relatively faint  bursts.

\section{Bursting Activity}
\subsection{Detection and Localization}
\begin{figure}[ht]
\centerline{\psfig{figure=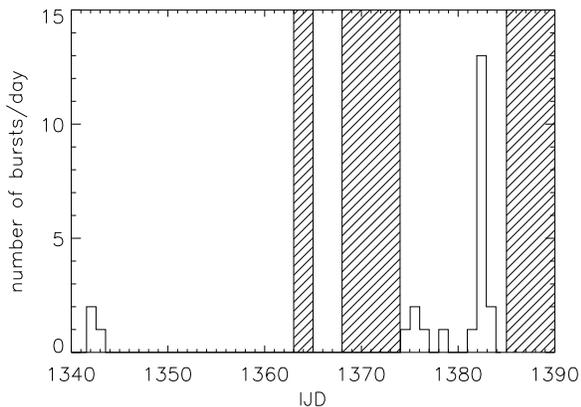,width=8.5cm}}
\caption{Burst detection rate per day. The shaded areas represent the time periods
when the source was outside the IBIS field of view (short gaps due to perigee passages 
and satellite slews are not indicated).}
\label{detection}
\end{figure}

In July 2003 a new active period of \src was detected by the IPN 
(see e.g. \citet{hurley2}). The source remained active throughout the
month of August and so an INTEGRAL Target of Opprotunity Observation
(TOO) was triggered. INTEGRAL observed it for 240 ks 
 starting on September 3 2003, and 3 faint bursts
were detected. Afterwards the monitoring of the source
continued during the Galactic Center Deep Exposure (GDCE) 
as part of INTEGRAL Core Program observations
(yielding an additional exposure of about $\sim$1 Ms on the source), during
which 21 bursts have been detected. All the bursts have been detected
and localized in near real time by the INTEGRAL Burst Alert System
(IBAS; \citet{ibas}), using IBIS \citep{ibis} data in the 15-200 keV range.
The burst detection rate vs. time is shown in Fig. \ref{detection}.

All the bursts have been localized within 2 arcminutes from the 
X-ray position of the quiescent counterpart \citep{chandra}. Since the 
typical error circle is 2.5 arcmin (90\% c.l., see \citet{ibas2}), 
we are confident that all the bursts
originate from \src and not from the possible newly discovered Soft Gamma-Ray 
Repeater SGR 1808-20 \citep{lamb}, which is located at 15 arcmin from the source.

\subsection{Data Analysis}

\begin{figure*}[ht!]
\centerline{\psfig{figure=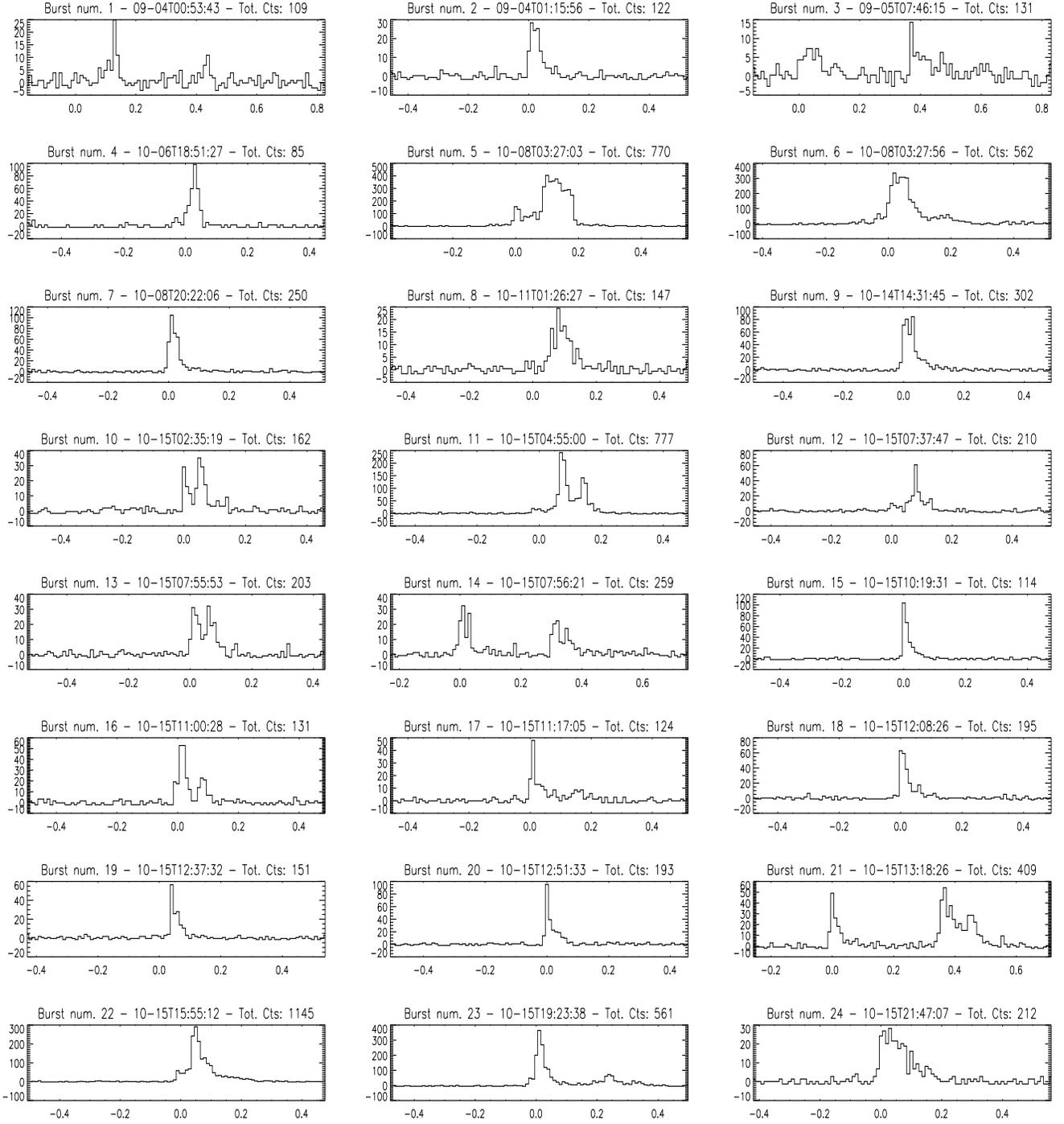,angle=90,width=18cm,height=19cm}}
\caption{IBIS/ISGRI background subtracted light curves of
the SGR 1806-20 bursts in the 15-100 keV range.
Each panel corresponds to a time interval of one second
and the time bins are of 10 ms.
Units of the axes are time in seconds and vignetting-corrected counts per bin.
Time=0 corresponds to the
the starting time of the $T_{90}$ computation and is 
reported on top of each panel (month - day - UT time)
together with the total number of net counts.}
\label{lc}
\end{figure*}

We have analyzed only ISGRI \citep{isgri} data, since 
PICsIT \citep{picsit} in its standard operation mode does not have the combined time resolution 
and sensitivity to study  such short bursts.  
The background subtracted IBIS/ISGRI light curves 
of the bursts, binned at 10 ms, are shown in Fig. \ref{lc}.
\begin{table*}
\begin{center}
\begin{tabular}{|c|c|c|c|c|c|c|}
\hline
Burst & Fluence & Peak Flux &  $T_{90}$ &Net& Off-Axis Angle Y& Off Axis-Angle Z\\
Number & 10$^{-8}$ erg cm$^{-2}$ & 10$^{-7}$ erg cm$^{-2}$ s$^{-1}$  & ms & Counts & degrees & degrees\\
\hline
       1 &       1.65 $\pm$     0.16 &       3.74 $\pm$     0.75 &      437 &      109& -3.89 & -1.86\\
       2 &       1.84 $\pm$     0.17 &       4.26 $\pm$     0.80 &       69 &      122& -3.96 & 0.15\\
       3 &       1.97 $\pm$     0.17 &       2.15 $\pm$     0.57 &      579 &      131& 4.05 & 0.17\\
       4 &       4.92 $\pm$     0.27 &       14.77 $\pm$    1.49 &      109 &       85& 10.26 & 8.21\\
       5 &       65.35 $\pm$    0.99 &       60.73 $\pm$    3.02 &      179 &      770& 3.36 & -12.96\\
       6 &       46.58 $\pm$    0.84 &       50.42 $\pm$    2.75 &      199 &      562& 3.35 & -12.92\\
       7 &       5.49 $\pm$     0.28 &       15.73 $\pm$    1.53 &       99 &      250& 3.64 & -7.56\\
       8 &       2.21 $\pm$     0.18 &       3.67 $\pm$     0.74 &      169 &      147& -0.97 & -2.22\\
       9 &       6.15 $\pm$     0.30 &       12.71 $\pm$    1.38 &       89 &      302& -6.32 & -4.49\\
      10 &       2.89 $\pm$     0.21 &       5.27 $\pm$     0.89 &      139 &      162& -5.61 & 0.77\\
      11 &       20.99 $\pm$    0.56 &       36.40 $\pm$    2.34 &      169 &      777& -0.71 & 8.87\\
      12 &       3.17 $\pm$     0.22 &       9.18 $\pm$     1.17 &      129 &      210& -3.69 & 3.77\\
      13 &       3.06 $\pm$     0.21 &       4.83 $\pm$     0.85 &       89 &      203& -3.69 & 3.73\\
      14 &       3.89 $\pm$     0.24 &       4.86 $\pm$     0.85 &      489 &      259& -3.69 & 3.78\\
      15 &       3.94 $\pm$     0.24 &       15.62 $\pm$    1.53 &       69 &      114& -10.03 & -3.26\\
      16 &       3.50 $\pm$     0.23 &       7.95 $\pm$     1.09 &       99 &      131& -8.63 & -1.40\\
      17 &       2.72 $\pm$     0.20 &       7.21 $\pm$     1.04 &      189 &      124& -7.29 & 0.50\\
      18 &       3.62 $\pm$     0.23 &       9.45 $\pm$     1.19 &       89 &      195& -5.95 & 2.44\\
      19 &       2.50 $\pm$     0.19 &       8.49 $\pm$     1.13 &      119 &      151& -4.66 & 4.42\\
      20 &       3.71 $\pm$     0.24 &       14.33 $\pm$    1.47 &      159 &      193& -3.45 & 6.41\\
      21 &       7.88 $\pm$     0.34 &       8.15 $\pm$     1.11 &      475 &      409& -3.46 & 6.42\\
      22 &       27.21 $\pm$     0.64 &       43.71 $\pm$   2.56 &      188 &     1145& -3.87 & 8.10\\
      23 &       25.17 $\pm$     0.61 &       54.80 $\pm$   2.87 &      288 &      561& -11.12 & -2.48\\
      24 &       4.29 $\pm$     0.25 &       4.24 $\pm$     0.80 &      159 &      212& -5.67 & 5.12\\
\hline
\end{tabular}
\end{center}
\caption{Fluences (15-100 keV), Peak Fluxes (over 10 ms, 15-100 keV), $T_{90}$ durations, net counts (background subtracted, but not corrected for vignetting) and instrumental coordinates for all the bursts.}
\label{bigtab2}
\end{table*}
In order to increase the signal-to-noise-ratio, they were extracted from ISGRI
pixels illuminated by the source for at least half of their surface
and selecting counts in the  15-100 keV energy range (most of the bursts
had little or no signal  at higher energy).

\begin{figure}[ht!]
\centerline{\psfig{figure=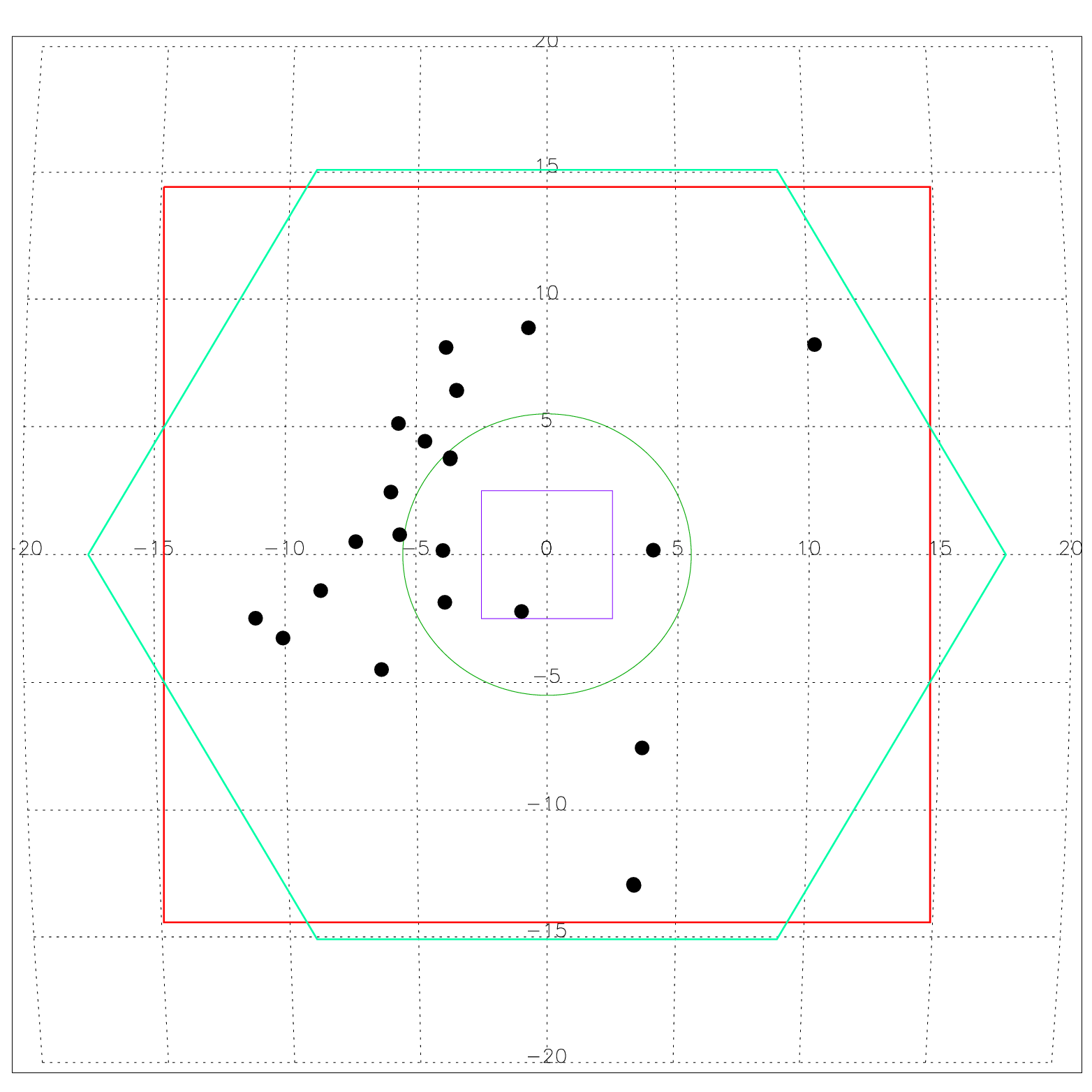,width=7.5cm}}
\caption{Burst positions in instrumental coordinates. 
The fields of view of IBIS (large square), SPI (hexagon), 
JEM-X (circle) and OMC (small
square) are plotted. The units of the grid are degrees.}
\label{fov}
\end{figure}

\begin{figure}[ht]
\centerline{\psfig{figure=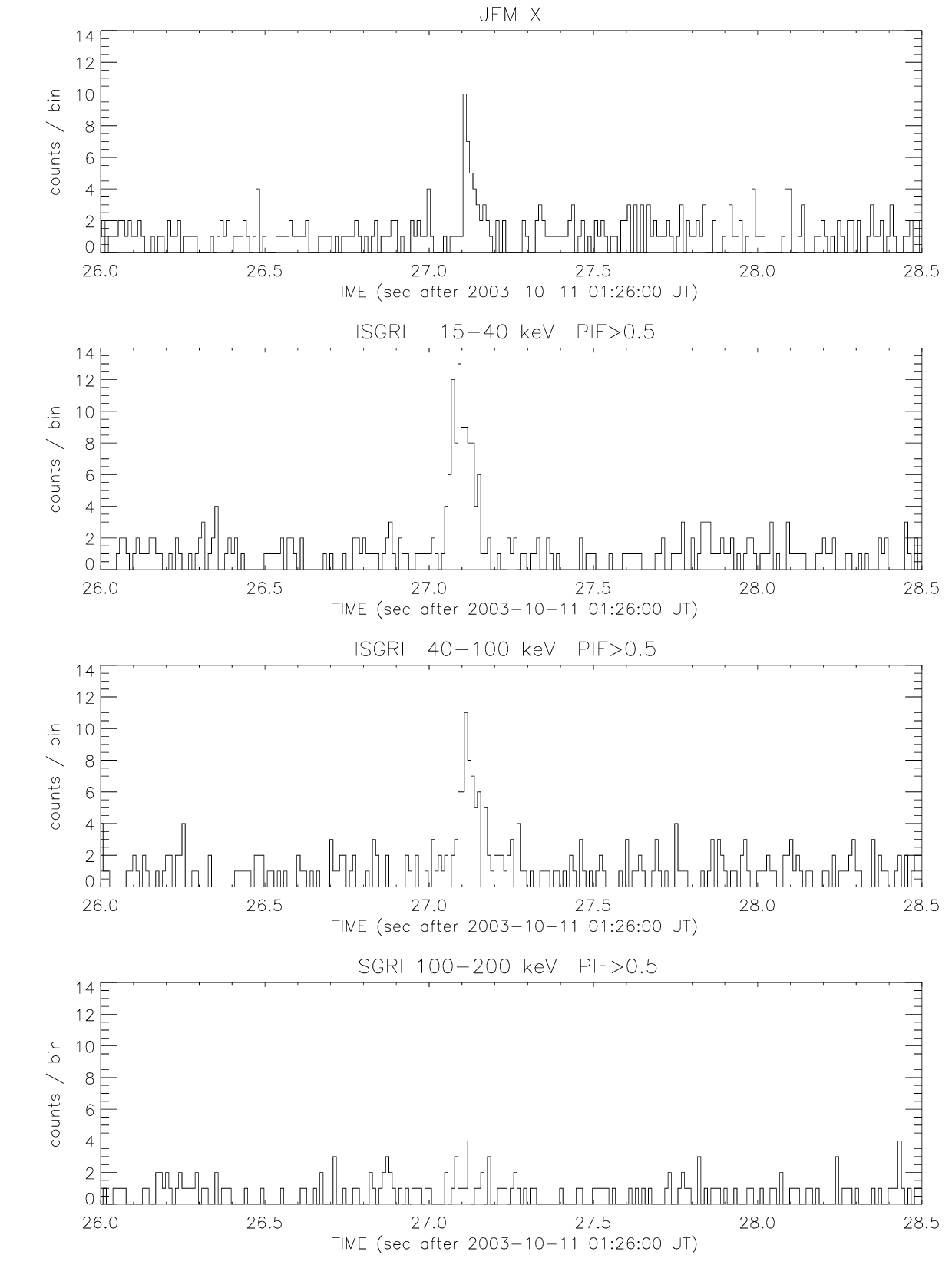,width=7.5cm}}
\caption{Light curves of the SGR burst detected also by JEM-X.}
\label{ibisjemx}
\end{figure}

The bursts were detected at various off-axis angles, ranging
from 2.5 to 13.3 degrees, corresponding to a variation of 80\% in 
the instrument effective area. The light curves shown 
in Fig. \ref{lc} have been corrected for this vignetting effect.  
The total number of net counts actually recorded for each burst
is indicated in the corresponding panel.
The burst positions in instrumental
coordinates are plotted in Fig. \ref{fov} (see also Tab. \ref{bigtab2}).
As can be seen most of the bursts are located in the partially coded field of
view of IBIS. A few bursts fall within the JEM-X \citep{jemx} 
field of view, but unfortunately
they are the faintest ones. The only burst for which a simultaneous IBIS/JEM-X 
detection has been obtained (see Fig. \ref{ibisjemx}), is the one with instrumental 
coordinates  Y=--0.97$^{\circ}$ and Z=--2.22$^{\circ}$ (burst number 8). This burst has a rather small fluence: 
2.2$\times$10$^{-8}$ erg cm$^{-2}$ (15-100 keV).

None of the bursts has been found in a preliminary analysis of SPI \citep{spi} 
data, but further analysis is underway. 
Only one burst fell into the OMC \citep{omc} 
field of view, but the large extinction towards the object
(A$_{v}\sim$ 30 mag) prevents the detection of any optical
emission (see also \citet{castro}).

The light curves shown in Fig. \ref{lc} have  shapes  typical for SGR bursts.
From the light curves we determined the $T_{90}$ duration of each burst 
(i.e. the time during which 90\% of the total burst counts are accumulated).
The $T_{90}$ values, reported in Tab. \ref{bigtab2},
range typically from $\sim$0.1 to $\sim$0.2 s for single
peaked bursts and can be as long as $\sim$0.6 s for double
peaked bursts. In fact the $T_{90}$ values of these bursts include
the ``interpulse'' period.
Some bursts are preceded by a small  precursor.

\begin{figure}[ht]
\centerline{\psfig{figure=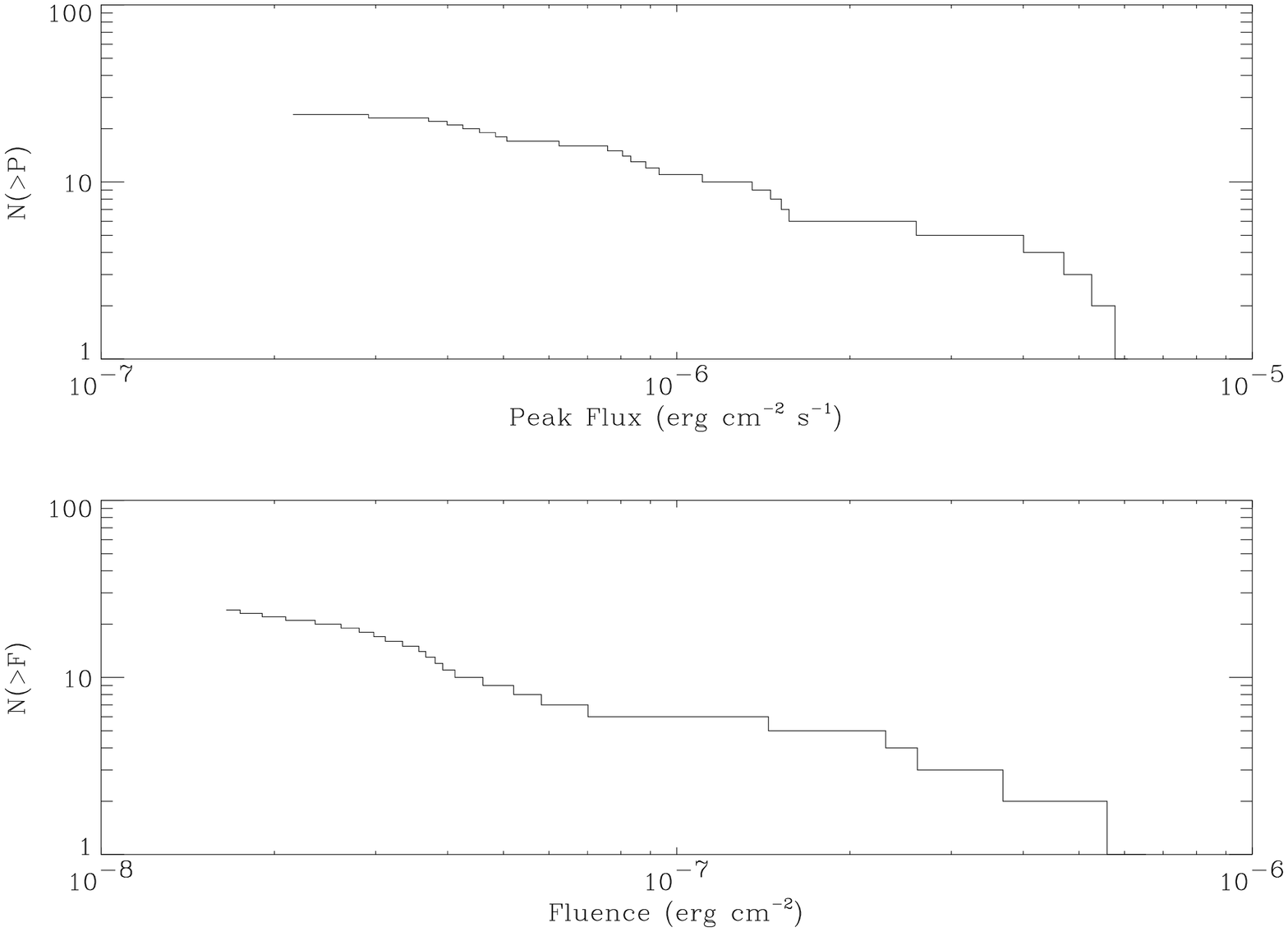,angle=90,width=7.5cm, height=10cm}}
\caption{Integral distributions of Peak flux
      ($\Delta t$=10 ms, \emph{Upper Panel}) and fluences (\emph{Lower Panel}).
      Both refer to the 15-100 keV energy range.}
\label{log}
\end{figure}

The peak flux  and fluence for each burst were first derived in counts units
from the light curves of Fig.\ref{lc}, and then converted to physical units
adopting a constant conversion factor derived from the spectral analysis of the
brightest bursts (see below).
The resulting 15-100 keV peak fluxes and fluences are reported in Tab.
\ref{bigtab2} and are respectively in the range
(4--50)$\times10^{-7}$ erg cm$^{-2}$ s$^{-1}$ ($\Delta t$=10 ms)
and (2--60)$\times10^{-8}$ erg cm$^{-2}$.
Their integral distributions are shown in Fig. \ref{log}.
Within the large uncertainties, the fluence distribution is consistent
with the power law slope found by \citet{gogus}
using {\it RXTE} data. 
Many of these bursts are the faintest
ever imaged from SGRs at these energies.

\begin{figure}[ht!]
\centerline{\psfig{figure=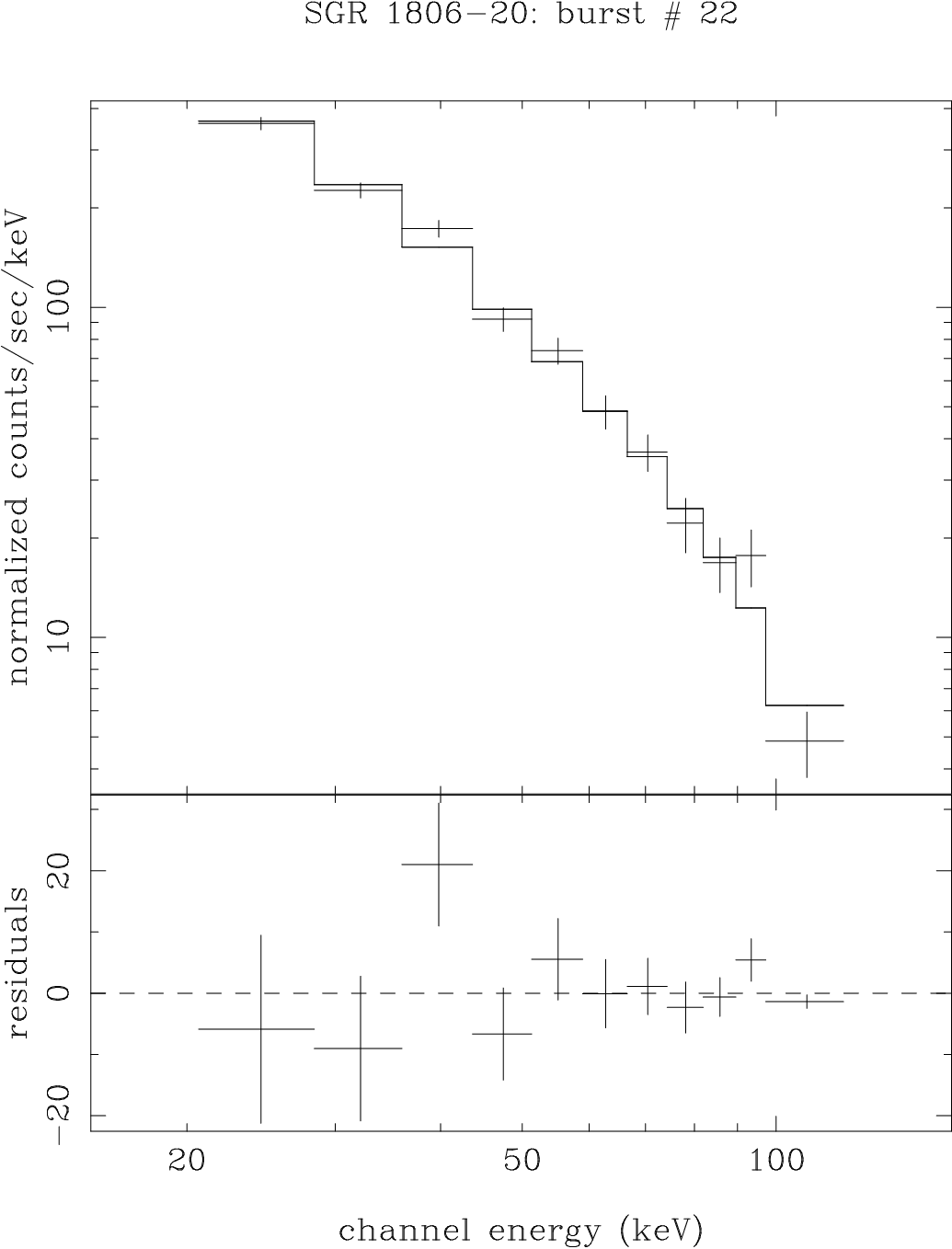,width=7.5cm}}
\caption{Time averaged spectrum of burst \# 22 (IBIS/ISGRI).}
         \label{sp}
 \end{figure}

For the  bursts with more than 500 net counts we could perform a detailed
spectral analysis.
The 15-200 keV spectra, integrated over the whole duration of each burst,
were well fitted by an Optically
Thin Thermal Bremsstrahlung (OTTB) model, which
is the spectral model that typically provides good
fits to SGR bursts above 20 keV.
Using the latest available response matrices,
we measured temperatures ($kT$) between 32 to 42 keV. These
values are in the typical range of temperatures found in the literature (see e.g. \citet{atteia})
for short bursts from \src.
We tried other  models, like a
power law or a black body, but they
were clearly ruled out.
Small known calibration uncertainties at low energies ($<$ 40 keV) do not affect
much our spectra, whose errors are dominated rather by statistical uncertainties
than by systematic ones.
The OTTB fit to the spectrum of burst number 22 is shown as an example in Fig. \ref{sp}.

Adopting a temperature $kT$=38 keV (consistent with the average
spectra of the brightest bursts) we derived a conversion factor of
1 count s$^{-1}$ = 1.5$\times$10$^{-10}$ erg cm$^{-2}$ s$^{-1}$
(15-100 keV), which we adopted for all the bursts.

\begin{figure*}[ht]
\centerline{\psfig{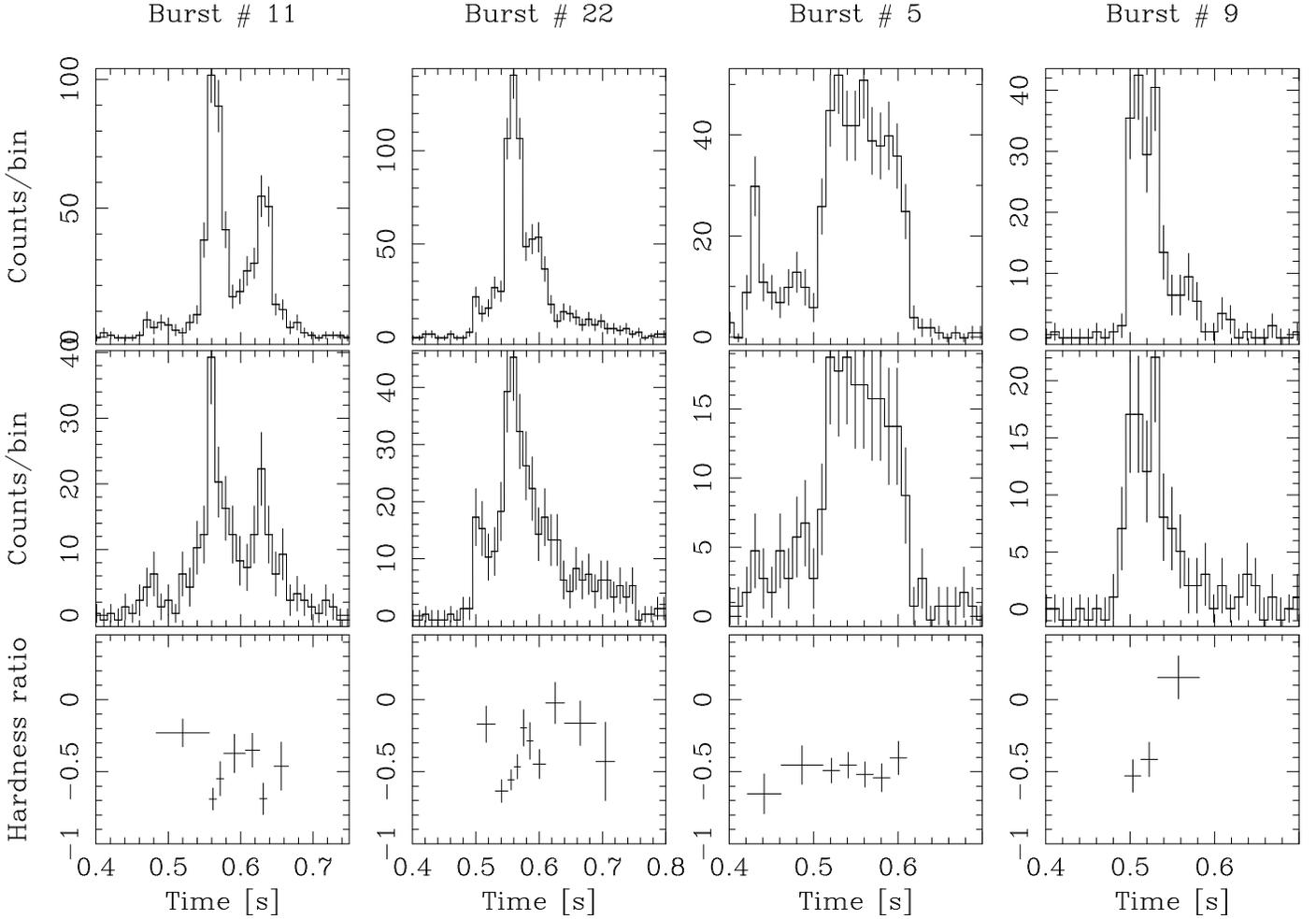}}
\caption{15-40 keV light curve (\emph{Top Panels}), 40-100 keV light curve
      (\emph{Middle Panels}), and time resolved hardness ratio (\emph{Bottom Panels})
      for four bursts with good statistics. The time resolved hardness ratio for
      bursts number 11,22,9 is inconsistent with a constant value at $\sim$3.5 
$\sigma$ level \citep{gotzc}.}
\label{trehr}
\end{figure*}

\begin{figure}[ht!]
\centerline{\psfig{figure=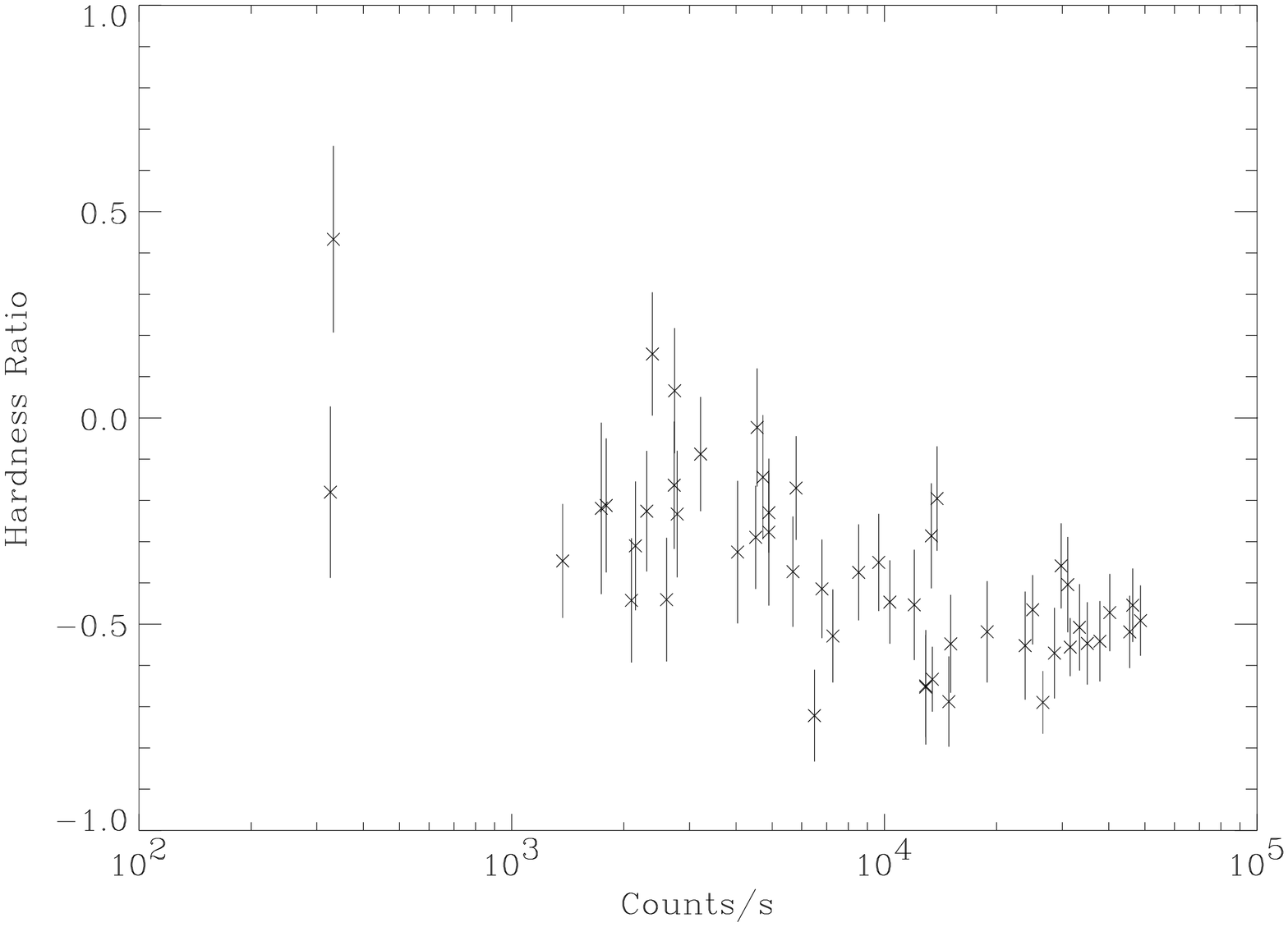,width=7.cm,angle=90}}
\caption{Hardness-Intensity plot of the time resolved hardness ratios of
the 12 bursts with the best statistics.
The hardness ratio is defined as $(H-S)/(H+S)$, where $H$ and $S$ are the
background subtracted counts in the ranges 40-100 keV and 15-40 keV respectively. The count rates are 
corrected for the vignetting and refer to the 15-100 keV range \citep{gotzc}.}
         \label{hi}
 \end{figure}

\citet{gotzc} have used these data, to investigate 
the time evolution of the bursts spectra,  
using time-resolved hardness ratios ($HR=(H-S)/(H+S)$, where $H$ and $S$ are the
background subtracted counts in the ranges 40-100 keV and 15-40 keV respectively).
Analyzing the 12 bursts with more than 200 net counts,
they found that some bursts show a significant spectral evolution, while
others, particularly those with a  ``flat topped'' profile, do not.
Some examples are given in  Fig. \ref{trehr}. 
In addition the variation of the hardness ratio
versus intensity ($I$) has been investigated. 
Considering all the time bins of all the bursts (see Fig. \ref{hi}),
\citet{gotzc} found a hardness-intensity anti-correlation. The linear correlation
coefficient of the data corresponds to a chance probalility 
smaller than 10$^{-3}$ of being due to uncorrelated data. In addition,
according to an F-test, the data are significantly (at a 5.2 $\sigma$ level) better
described by a linear fit than by a constant value.

We have verified that there is no correlation between the hardness ratios 
of the individual bursts and the off-axis angle at which they have been detected.
In adition we have verified that the vignetting correction procedure used in 
\citet{gotzc} is consistent with the values of the flux of the
Crab Nebula measured in different positions in the field of view of IBIS.

\section{Persistent Emission}
Persistent (quiescent) emission from \src 
has been discorvered at X-ray ($<$10 keV) energies \citep{asca}. 
Up to now no detection at higher energies has been reported for any of the SGRs.
We have analyzed IBIS, JEM-X and SPI data 
of the TOO (240 ksec), but we did not find convincing evidence of quiescent emission.

\begin{figure}[ht!]
\centerline{\psfig{figure=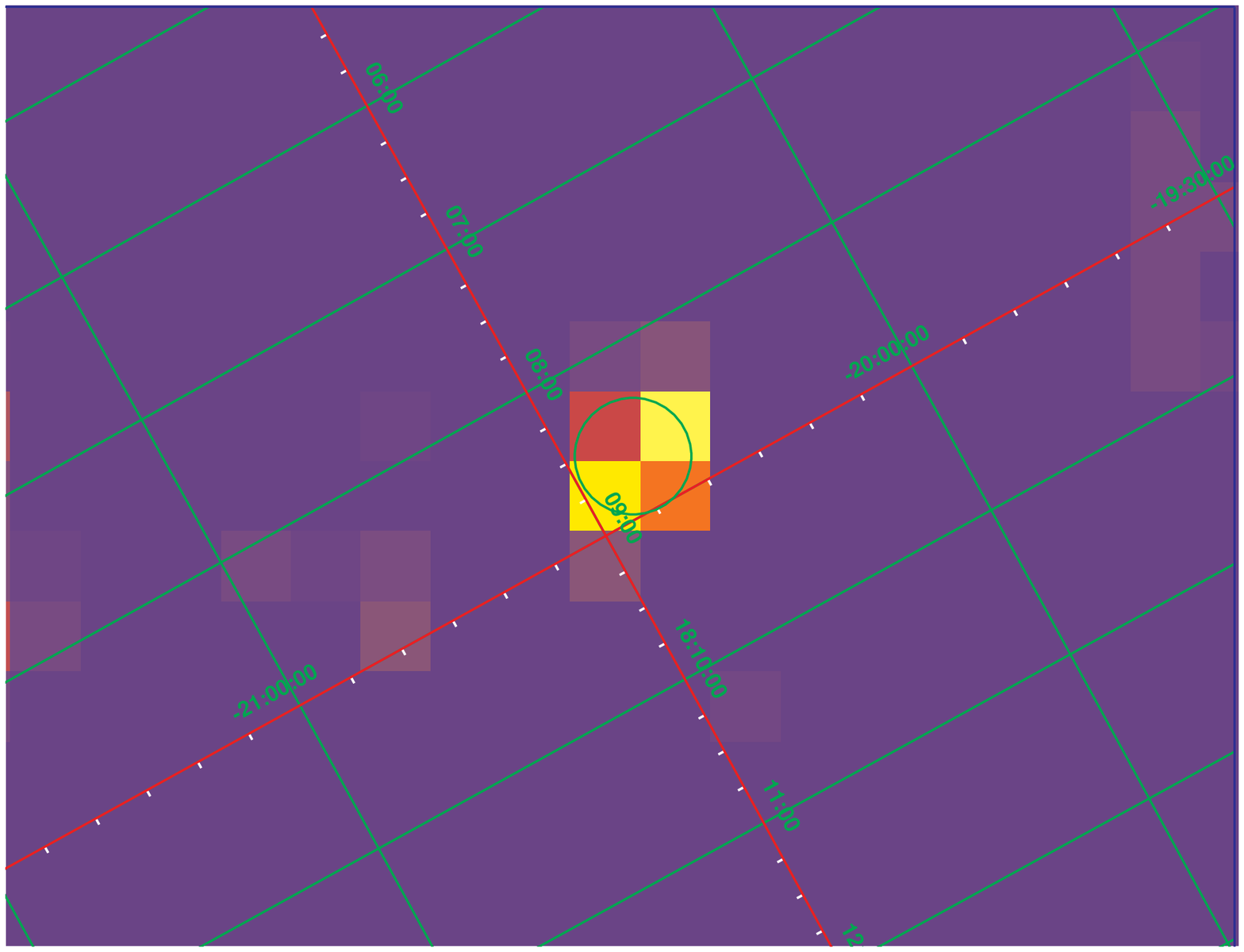,width=7.cm}}
\caption{IBIS/ISGRI significance map. \src is detected at $\sim$6 $\sigma$ level. 
The circle is centered on the X-ray Chandra position \citep{chandra} and has 5 arcmin radius.}
         \label{isgri}
 \end{figure}

On the other hand analyzing IBIS/ISGRI Core Program data ($\sim$1 Msec of exposure on \src),
we detect the source at $\sim$6 $\sigma$ level 
in IBIS/ISGRI (see Fig. \ref{isgri}) in the 20-40 keV band.
The flux is $\sim$3 mCrab, consistent with an extrapolation
of the spectrum measured at lower energies \citep{meresax}. Although a detection
with such a relatively small significance is also compatible with background 
systematic noise in the significance maps,
the positional coincidence with the X-ray counterpart strengthens the detection.

Also SPI data have been searched for quiescent emission.
482 pointings around the position of \src (Core Program and  TOO data) have been used, 
yielding a total exposure of $\sim$0.798 Ms. For the
SPIROS \citep{spiros} analysis tool the catalogue of the sources detected by IBIS/ISGRI
has been used as an input. A
significance map was generated in the energy range between 28 and 48 keV.
The measured flux is 0.69$\pm$0.15$\times$10$^{-3}$ photons cm$^{-2}$ s$^{-1}$.
This flux ($\sim$ 10 mCrab) is slighly higher than the one measured by IBIS,
but considering that the SPI detection level is just $\sim$ 4.4 $\sigma$, and that
systematic errors are still present at this stage in the analysis of both instruments,
we can consider the two results as consistent with each other.

Further GCDE observations during INTEGRAL AO-2 will surely help to further
assess the detection and the source flux.

Concerning JEM-X Core Program data, the analysis has still to be 
completed and will surely be helpful in order to better asses the possible IBIS
and SPI detection.

The detection of pulsed emission is at the moment below our sensitivity.

\section{Conclusions}

We can summarize our results as follows:
\begin{itemize}
\item For the first time good evidence for spectral evolution
of weak SGR bursts is presented and a hardness-intensisty
anti-correlation within the bursts has been found \citep{gotzc}. This results
represent a new challenge for the \emph{Magnetar} model, which
predicts that the effetive temperature of the burst should vary weakly
during the bursts, while we detect large spectral variations whitin
the bursts as for number 9 and 22.

\item We report for the first time the possible detection, in IBIS and SPI data,
of the quiescent emission above 10 keV.

\item IBIS is a very sensitive detector for SGR bursts, with fluences down to 
$\sim$10$^{-8}$ erg cm$^{-2}$ in imaging mode. In fact in just 1.5 months it 
detected twice as many bursts from this source as BATSE during its whole lifetime \citep{gogus}.

\item The search of pulsed emission is at the moment below 
the sensitivity threshold for our data.

\item \src will be observed again during INTEGRAL AO-2 allowing
to study this source more deeply. 

\end{itemize}

We finally note that since January 2004 the INTEGRAL Burst Alert System\footnote{http://isdc.unige.ch/index.cgi?Soft+ibas} 
(IBAS) is distributing alerts also for SGR Bursts. 
Two examples are a $\sim$0.2 s long burst from \src detected at
10:24:08.35 UT on March 9 2004 \citep{gotzd} and the $\sim$0.5 s long faint burst
detected at 13:02:06.33 UT on March 16 (IBAS Alert n. 1677).
The alert messages have been distributed
just $\sim$20 seconds after the bursts, allowing for prompt follow-up observations at other wavelenghts.
This feature will be particularly useful in case of major bursts 
like the ones from SGR 1900+14 or SGR 0526-66.

\section*{Acknowledgmentes}
This work has been partially supported by the Italian Space Agency (ASI). 
DG is grateful to Ada Paizis for analyzing IBIS Core Program data.

\end{document}